\documentclass[aps, prl,10pt,notitlepage,nofootinbib,superscriptaddress,showkeys,showpacs,twocolumn]{revtex4}
\usepackage{amstext,amsmath,amssymb,amsfonts,bbm, amsthm}
\usepackage[latin1]{inputenc}
\usepackage{fancyhdr}
\usepackage{graphicx}
\usepackage{hyperref}
\usepackage{tocvsec2}
\usepackage{cleveref}

\usepackage{color}

\def\beq{\begin{equation}}
\def\be{\begin{equation}}
\def\ee{\end{equation}}
\def\bes{\begin{eqnarray}}
\def\ees{\end{eqnarray}}

\def\f{\frac}

\def\pp{\partial}

\begin{document}

\title{\large \bf Stochastic Thermodynamics of Entropic Transport}


\author{{Matteo Smerlak}}\email{smerlak@aei.mpg.de}
\affiliation{Max-Planck-Institut f\"ur Gravitationsphysik, Am M\"uhlenberg 1, D-14476 Golm, Germany}
\date{\small\today}

\begin{abstract}\noindent
Seifert derived an exact fluctuation relation for diffusion processes using the concept of ``stochastic system entropy''. In this paper we extend his formalism to entropic transport. We introduce the notion of \emph{relative stochastic entropy} and use it to generalize Seifert's system/medium decomposition of the total entropy. This result allows to apply the concepts of stochastic thermodynamics to diffusion processes in confined geometries, such as ion channels, cellular pores or nanoporous materials. It can be seen as the equivalent for diffusion processes of Esposito's and Schaller's generalized fluctuation theorem for ``Maxwell demon feedbacks''. 
\end{abstract}
\maketitle

From a mathematical perspective, the fluctuation relations of non-equilibrium statistical mechanics \cite{Jarzynski2011} boil down to a couple of---almost-tautological---facts: 
\begin{itemize}
\item
If $\omega$ is a random variable, $P(\omega)$ and $P'(\omega)$ are two probability distributions on $\omega$ and
\be\label{defR}
R(\omega)=\ln \f{P(\omega)}{P'(\omega)},
\ee
then the normalization of $P'$ implies
\be\label{IFR}
\langle e^{-R}\rangle\equiv\sum_{\omega}P(\omega)\,e^{-R(\omega)}=1,
\ee 
and therefore, by Jensen's inequality, 
\be\label{2L}
\langle R\rangle\geq0.
\ee
\item
In the particular case where $P'(\omega)=P(\omega^{\dagger})$ for some involution $(\,\cdot\,)^{\dagger}$ (viz. $\omega^{\dagger\dagger}=\omega$), then we have the stronger relation
\be\label{DFR}
\textrm{Prob}\big(R(\omega)=-r\big)=e^{-r}\,\textrm{Prob}\big(R(\omega)=r\big).
\ee
\end{itemize}

As stressed by Maes \cite{Maes1999,Maes2000,Maes2003}, the relevance of these lemmas for non-equilibrium statistical mechanics becomes apparent relevant when $(i)$ $\omega$ represents a path in the (suitably coarse-grained) state space of a mesoscopic system and $(ii)$ $(\,\cdot\,)^{\dagger}$ is the time-reversal operation. In this case, indeed, the quantity $R(\omega)$ often turns out to have a \emph{thermodynamic interpretation}, in terms of the work received by the system, the heat dissipated or the entropy produced along $\omega$. The identity \eqref{DFR} then becomes a Gallavotti-Cohen \cite{Gallavotti1995} type \emph{detailed fluctuation relation} (DFR), the more general identity \eqref{IFR} becomes a Jarzynski \cite{Jarzynski1997} type \emph{integral fluctuation relation} (IFR), and the inequality \eqref{2L} becomes the second law inequality. This unifying perspective on the non-equilibrium fluctuation relations is developed in \cite{Chetrite2007,Seifert2012}.

In the context of overdamped Brownian motion, where $\omega=(x_{t})_{0\leq t\leq T}$ denotes the stochastic trajectory of a particle driven by a protocol-dependent force $F(x,\lambda)$ (in contact with a bath at inverse temperature $\beta$), this thermodynamic interpretation was clarified by Seifert in \cite{Seifert2005a}: due to the Markovian nature of Brownian motion, the irreversibility function $R(\omega)$ splits into a ``boundary'' term
\be
\Delta s(\omega)=-\ln p(x_{T},T)+\ln p(x_{0},0),
\ee
where $p(x,t)$ is the solution of the corresponding Fokker-Planck equation with initial condition $p(x,0)$, and a ``bulk'' term\footnote{The stochastic integral should be interpreted in the Stratonovitch sense.}
\be
\beta Q(\omega)=\beta\int_{0}^{T}F(x_{t},\lambda_{t})\cdot dx_{t}. 
\ee 
The interpretations of these terms are, respectively, the \emph{stochastic system entropy}---we prefer the term ``surprisal'' \cite{Tribus1961}---variation, and ($\beta$ times) the \emph{dissipated heat} along $\omega$. In other words, $R(\omega)$ is nothing but the \emph{total entropy variation}
\be\label{seifert}
R(\omega)=\Delta s_{\textrm{tot}}(\omega)=\Delta s(\omega)+\beta Q(\omega). 
\ee
The identities \eqref{IFR}, and \eqref{DFR} in the case of steady states, provide fluctuation relations for the total entropy production. This interpretation forms the backbone of \emph{stochastic thermodynamics} \cite{Seifert2008,Seifert2012}. 

The purpose of this note is to point out that Seifert's interpretation \eqref{seifert} must be generalized to become applicable for \emph{entropic transport}, that is, to diffusion processes where the density (or degeneracy)  of states $\Omega(x)$---hence the hence the free equilibrium state $p^{*}(x)\propto\Omega(x)$---is state-dependent. An example of such situation is the ``entropic barrier'' studied by Jacobs \cite{Jacobs1967} and Zwanzig \cite{Zwanzig1992}: a narrow channel whose sectional area $A(z)=\pi R^{2}(z)$ varies along the main axis $z$. This entropic barrier models various physically, chemically and biologically relevant transport processes, including tracer diffusion accross ion channels, biological membranes, zeolites and nano-porous materials; see \cite{Burada2009} for a review. Provided the transverse dimensions are small enough for the corresponding diffusion times to be negligible, and assuming that $\vert R'(z)\vert\lesssim1$, the Fokker-Planck equation for the longitudinal motion (in the absence of forcing) is the Fick-Jacobs equation \cite{Jacobs1967,Zwanzig1992}
\be
\pp_{t}p(z,t)=\pp_{z}\left(D(z)A(z)\pp_{z}\Big(\f{p(z,t)}{A(z)}\Big)\right).
\ee
where $D(z)$ is an effective state-dependent diffusivity. The density of states of this projected diffusion process is $\Omega(z)=A(z)$, and the equilibrium state takes the form $p^{*}(z)\propto A(z)$.\footnote{Such non-standard Fokker-Planck equations, with space-dependent density of states, arise in other contexts as well, including general relativity \cite{Smerlak2011b}. In this case, the role of the spatially-varying density of states $\Omega(x)$ is played by the inverse of the comoving lapse function.} More generally, entropic transport phenomena are modeled by general Fokker-Planck equation of the form 
\be\label{FP}
\pp_{t}p(x,t)=\nabla\cdot\left(\mu(x)F(x,\lambda)p(x,t)+D(x)\nabla\Big(\f{p(x,t)}{\Omega(x)}\Big)\right),
\ee
with the local Einstein relation $\mu(x)=\beta D(x)$.

It is easy to see that Seifert's intepretation \eqref{seifert} does not hold for these state-dependent diffusion processes. Consider for simplicity the forcing-free case, $F=0$ in \eqref{FP}. Because the conventional (``Gibbs'') entropy is maximized by uniform distributions, and $p^{*}(x)\neq\textrm{const}.$, it suffices to take $p(x_{0},0)=\textrm{const}$ to find
\be
\langle\Delta s_{\textrm{tot}} \rangle =\langle \Delta s\rangle=\Delta \left(-\int dx\, p(x,t)\ln p(x,t)\right)<0,
\ee 
in direct violation of \eqref{2L}. This shows that $R(\omega)\neq\Delta s_{\textrm{tot}}(\omega)$ in this case, i.e. that the irreversibility of entropic transport is \emph{not} measured by the combination of the ``surprisal'' $-\ln p(x_{t},t)$ and the dissipated heat $Q(\omega)$, as in Seifert's relation \eqref{seifert}. In Bayesian terms, this is because the non-trivial density of states $\Omega(x)$ is a relevant \emph{prior}, which must be taken into account in the very definition of the surprisal. 

Taking our cues from Kullback and Leibler's definition of \emph{relative entropy} \cite{Kullback1959}, 
\be
D_{\textrm{KL}}[p\vert\Omega]=-\int dx\, p(x,t)\ln\f{p(x,t)}{\Omega(x)},
\ee
which is well-known to satisfy $\dot{D}_{\textrm{KL}}[p\vert\Omega]\geq0$ for any solution $p(x,t)$ of the Fokker-Planck equation \eqref{FP} with $F=0$, let us define the \emph{relative system entropy variation} or \emph{relative surprisal variation} by 
\be
\Delta s_{\textrm{rel}}(\omega,t\vert\Omega) \equiv-\ln\f{p(x_{T},T)}{\Omega(x_{T})}+\ln\f{p(x_{0},0)}{\Omega(x_{0})}. 
\ee
Here as before, $p(x,t)$ is the solution of the Fokker-Planck equation with initial condition $p(x,0)$. The Bayesian interpretation of the relative surprisal $s_{\textrm{rel}}(\omega,t\vert\Omega)\equiv-\ln p(x_{t},t)/\Omega(x_{t})$ is the following: $s_{\textrm{rel}}(\omega,t\vert\Omega)$ is the surprise experienced by an observer finding the system in state $x_{t}$ at time $t$ when she expected to find it in the forcing-free equilibrium state $p^{*}(x)\propto\Omega(x)$. 

Now the main point: for a diffusion process described by the Fokker-Planck equation \eqref{FP}, consider the probability $P[\omega]$ of a path $\omega=(x_{t})_{0\leq t\leq T}$ under the driving $\lambda=(\lambda_{t})_{0\leq t\leq T}$, and the probability $P'[\omega]\equiv\mathcal{D}P^{\dagger}[\omega^{\dagger}]$ of the time-reversed path $\omega^{\dagger}\equiv(x_{T-t})_{0\leq t\leq T}$ under the time-reversed protocol $\lambda^{\dagger}\equiv(\lambda_{T-t})_{0\leq t\leq T}$. Then the irreversibility function $R(\omega)$ defined by \eqref{defR} decomposes as 
\be\label{result}
R(\omega)=\Delta s_{\textrm{rel}}(\omega\vert\Omega)+\beta Q(\omega). 
\ee
To show this, first isolate the initial and distributions $p(\cdot,0)$ and $p(\cdot,T)$,
\be\label{isol}
R(\omega)=\ln\f{p(y_{0},0)}{p(y_{1},T)}+\ln \f{P[\omega\vert  x_{0}=y_{0}]}{P^{\dagger}[\omega^{\dagger}\vert x_{T}=y_{1}]}
\ee
and, second, use the Girsanov formula as in \cite{Lebowitz1999}, 
\begin{multline}
\ln \f{P[\omega\vert  x_{0}=y_{0}]}{P^{\dagger}[\omega^{\dagger}\vert x_{T}=y_{1}]}
=\\\int_{0}^{T}\f{\mu(x_{t})F(x_{t},\lambda_{t})-D(x_{t})\nabla \Omega^{-1}(x_{t})}{D(x_{t})\Omega^{-1}(x_t)}\cdot\,dx_{t}.
\end{multline}
With the Einstein relation $\beta D(x)=\mu(x)$, this gives
\be
\ln \f{\mathcal{D}P[\omega\vert  x_{0}=y_{0}]}{\mathcal{D}P[\omega^{\dagger}\vert x_{T}=y_{1}]}=\beta Q(\omega)-\int_{0}^{T}\f{\nabla \Omega^{-1}(x_t)}{\Omega^{-1}(x_t)}\cdot\, dx_{t},
\ee
and evaluating the second integral explicitly, we arrive at 
\be\label{genDFT}
\ln \f{\mathcal{D}P[\omega\vert  x_{0}=y_{0}]}{\mathcal{D}P^{\dagger}[\omega^{\dagger}\vert x_{T}=y_{1}]}=\beta Q(\omega)+\ln\f{\Omega(y_{1})}{ \Omega(y_{0})}.
\ee
Combining \eqref{genDFT} with \eqref{isol} gives the desired result \eqref{result}. 

Accordingly, the IFR for entropic transport can be written
\be
\langle e^{-\Delta s_{\textrm{rel}}(\omega\vert\Omega)+\beta Q(\omega)}\rangle=1.
\ee 
In the special case of steady states, $p(\cdot,0)=p(\cdot,T)$, with steady forcing $\lambda_{T-t}=\lambda_{t}$---in which case $P^{\dagger}=P$---we have the corresponding DFR  
\begin{multline}
\textrm{Prob}\big(\Delta s_{\textrm{rel}}(\omega\vert\Omega)-\beta Q(\omega)=-r\big)\\=e^{-r}\,\textrm{Prob}\big(\Delta s_{\textrm{rel}}(\omega\vert\Omega)-\beta Q(\omega)=r\big).
\end{multline}
The second law inequality, in turn, reads
\be
D_{\textrm{KL}}[p(\,\cdot\,,T)\vert\Omega]-D_{\textrm{KL}}[p(\,\cdot\,,0)\vert\Omega]\geq\beta\langle Q\rangle. 
\ee
These relations extend the framework of stochastic thermodynamics to entropic transport. 

We close by observing that our results are somewhat complementary to those of \cite{Esposito2012}. In that paper, Esposito and Schaller study, within the framework of Markov jump processes, the stochastic thermodynamics of ``Maxwell demon feedbacks''. These are defined as systems satisfying a modified detailed balance condition of the form
\be
\ln \f{W_{mm'}}{W_{m'm}}=-\beta\Big((\epsilon-\epsilon_{m})-\mu(N_{m}-N_{m'})\Big)+f_{mm'},
\ee
where $m$ is a state, $\epsilon_{m}$ its energy, $N_{m}$ its number of particles, $\mu$ the chemical potential, $W_{mm'}$ the $m\rightarrow m'$ transition rate, and $f_{mm'}$ a ``feedback parameter'' such that $f_{mm}=0$ and $f_{mm'}=-f_{m'm}$, measuring the departure for the standard detailed balance condition. Comparing with \eqref{genDFT}, we see that the feedback parameters play the same role in Esposito's and Schaller's setting as a non-trivial density of states $\Omega(x)$ in ours. This is intuitively clear: a gradient $\nabla\Omega(x)\neq0$ effectively acts as a ``Maxwell demon'' for the diffusing particles, biasing their Brownian motion towards regions of high $\Omega(x)$. As emphasized in \cite{Esposito2012}, the term $\ln\Omega(x_{T})/\Omega(x_{0})$ in \eqref{genDFT} can be interpreted as the integral of an ``information current''. 



\bigskip
The author thanks C. Jarzynski for bibliographical advice, R. Chetrite, M. Esposito and G. Schaller for useful comments a previous version of this manuscript, the physics department of Ecole normale sup\'erieure de Lyon (France) for an invitation to give a seminar on this topic, and the Inter-University Center for Astrophysics and Astronomy (India), where part of this work was performed, for hospitality.

\bibliographystyle{utcaps}
\bibliography{library}

\end{document}